\documentclass[3p,times,procedia]{elsarticle}
\usepackage{nupha_ecrc}
\usepackage{amsmath}

\volume{00}

\firstpage{1}

\journalname{Nuclear Physics A}

\runauth{}

\jid{nupha}

\jnltitlelogo{Nuclear Physics A}

\usepackage{amssymb}

\usepackage[figuresright]{rotating}

\begin{document}

\begin{frontmatter}



\dochead{XXVIIth International Conference on Ultrarelativistic Nucleus-Nucleus Collisions\\ (Quark Matter 2018)}

\title{Searching for a CEP signal with lattice QCD simulations}


\author[l1,l2,l3]{Zolt\'an Fodor}
\author[l3]{Matteo Giordano}
\author[l4]{Jana N. Guenther}
\author[l3]{Korn\'el Kap\'as}
\author[l3]{S\'andor D.\ Katz}
\author[l1]{Attila P\'asztor\footnote{speaker}}
\author[l5]{Israel Portillo}
\author[l5]{Claudia Ratti}
\author[l1,l2]{D\'enes Sexty}
\author[l2]{K\'alman K. Szab\'o}

\address[l1]{Department of Physics, University of Wuppertal, D-42119 Wuppertal, Germany}
\address[l2]{J\"ulich Supercomputing Centre, Forschungszentrum J\"ulich, D-52428 J\"ulich, Germany}
\address[l3]{Institute for Theoretical Physics, E\"otv\"os University, H-1117 Budapest, Hungary}
\address[l4]{Department of Physics, University of Regensburg, Universit\"atsstraße 31, 93053 Regensburg, Germany}
\address[l5]{Department of Physics, University of Houston, Houston, TX 77204, USA}

\begin{abstract}
We discuss the reliability of available methods 
to constrain the location of 
the QCD critical endpoint with lattice simulations. 
In particular we calculate the baryon fluctuations up 
to $\chi^B_8$ using simulations at imaginary chemical 
potentials. We argue that they contain no hint of 
criticality.
\end{abstract}


\end{frontmatter}


\section{Introduction: Lee-Yang zeros and a toy-model study at $N_t=4$}
The equation of state of QCD matter at $\mu=0$ has been calculated with lattice QCD 
techniques~\cite{Borsanyi:2013bia, Bazavov:2014pvz, Borsanyi:2016ksw}.
Calculating the equation of state also for nonzero density is a difficult task, because of the sign
and overlap problems. At the moment the state-of-the-art for realistic lattices is to 
calculate Taylor coefficients of the pressure
\cite{Borsanyi:2012cr, DElia:2016jqh, Bazavov:2017dus, Gunther:2016vcp, Borsanyi:2018grb}
either by direct calculations at $\mu=0$
or via fitting a polynomial ansatz to lattice simulation results at zero and imaginary chemical potentials.
There exist now lattice calculations of higher order fluctuations $\chi^B_6$ and $\chi^B_8$ for fine 
lattices, though the statistical errorbars on $\chi^B_8$ are still quite large. It is a natural and
important question to ask, whether these recent lattice results show any hints of criticality.
In this conference contribution we will examine this question in some detail.

The grand canonical partition function at finite volume is a finite order polynomial in $e^{\mu_q/T}$:
\begin{equation}
    \mathcal{Z}(\mu_q^2,T) = \sum_{N=-N_s^3}^{+N_s^3} Z_N(T) e^{3 N {\mu_q/T}} \rm{,}
\end{equation}
and its roots are called the Lee-Yang zeros.
It is also an even function in $\mu_q$, making the natural variable $\mu_q^2$.
The infinite volume limit of the Lee-Yang zeros gives important information about the thermodynamics 
of a system. Here we concentrate on the Lee-Yang zero closest to $\mu_q^2=0$, and call it $\mu^2_{LY}$.
If we have a nonzero infinite volume limit of $\operatorname{Im} \mu_{\rm{LY}}^2$ we have an analytic 
crossover. In this case $\operatorname{Im} \mu_{\rm{LY}}^2$ can be thought of as a measure of the 
strength of the crossover. For a genuine phase transition, we expect the infinite volume limit of
the Lee-Yang zero to be real. For any finite volume, the convergence radius of the Taylor 
expansion of the pressure in $\mu_q^2$ is given by the distance 
from $mu_q=0$ of the smallest Lee-Yang zero.

\begin{figure}[h]
\begin{center}
    \includegraphics[width=0.36\linewidth]{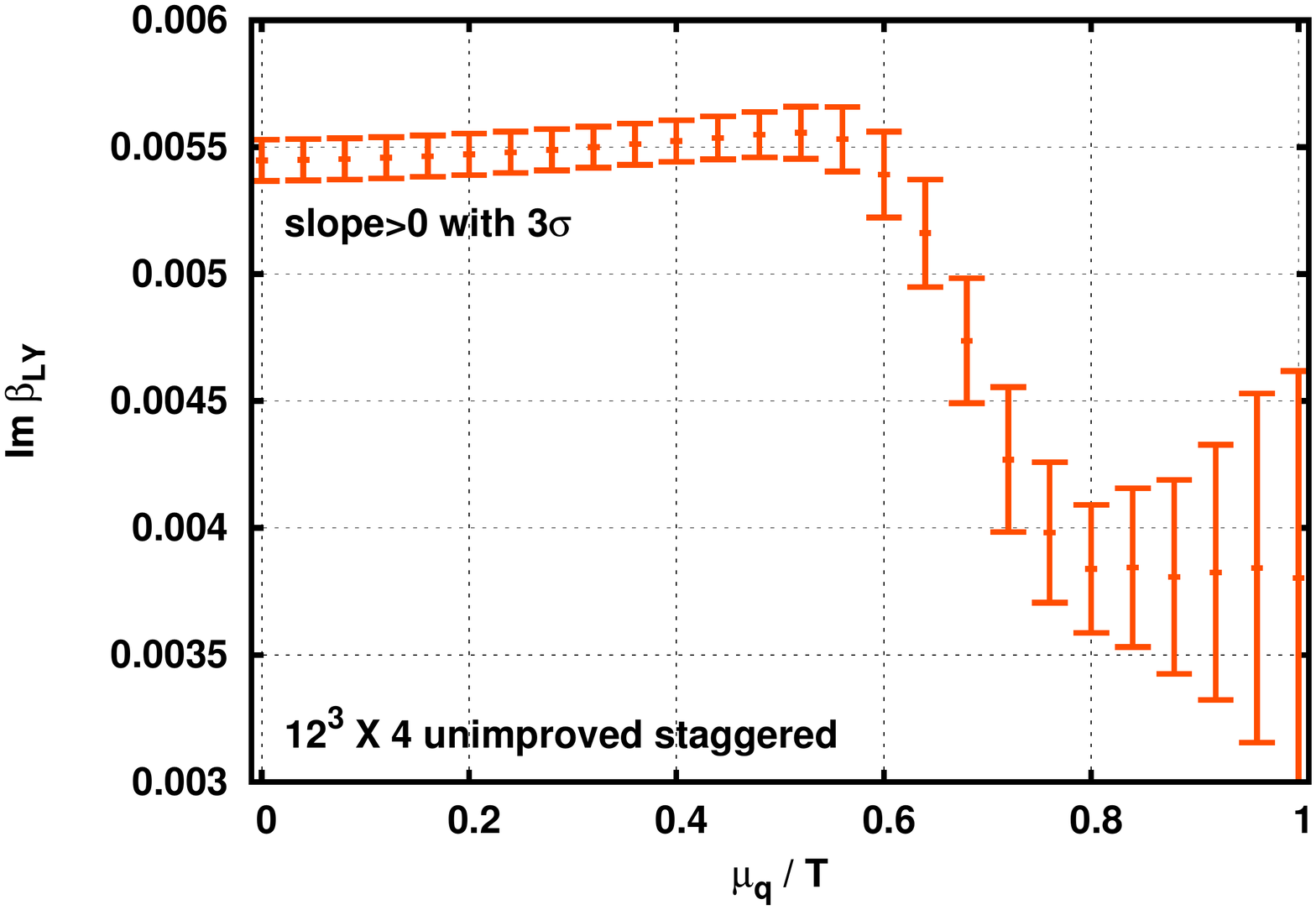}
    \hspace{-1.02cm}
    \includegraphics[width=0.36\linewidth]{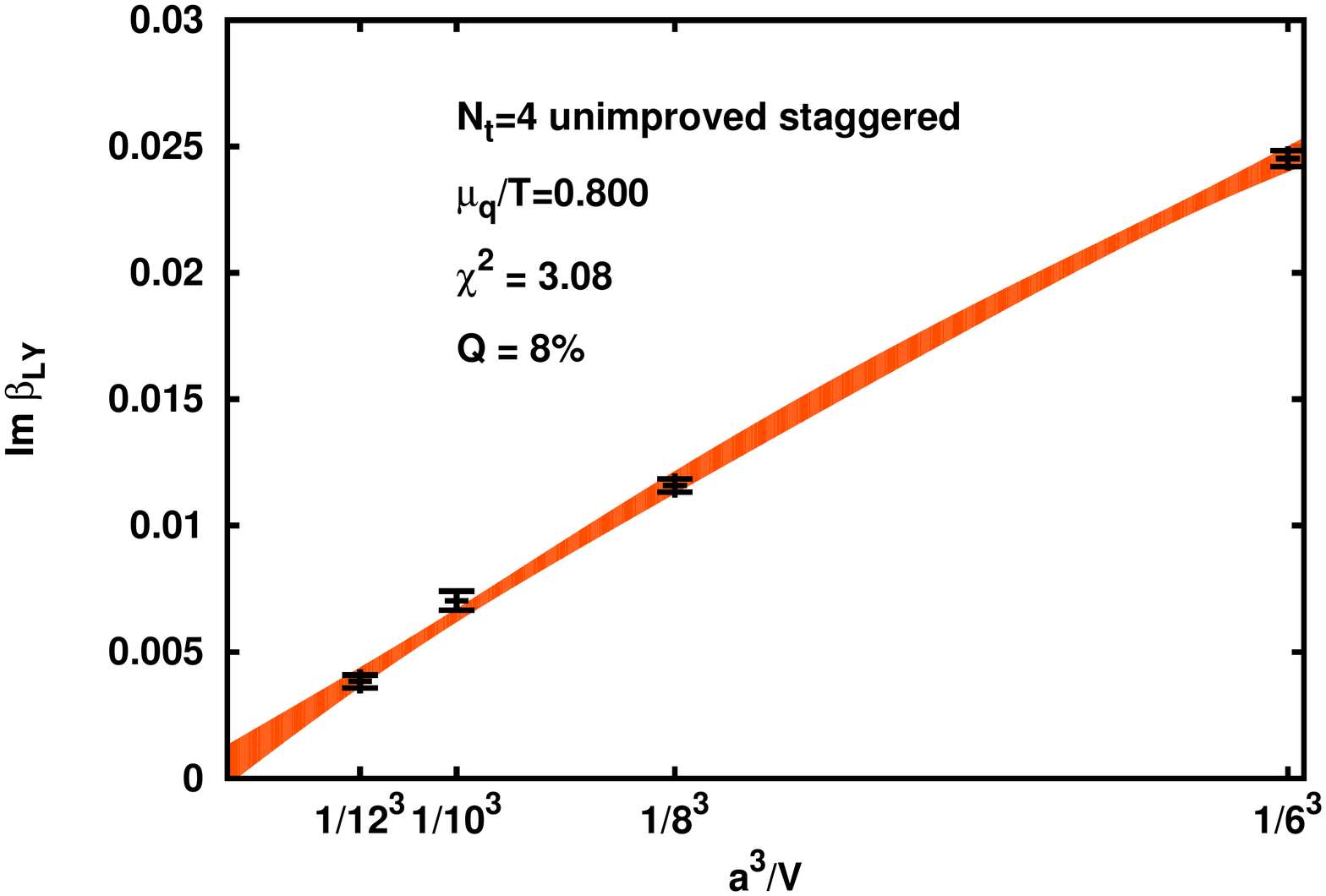}
    \hspace{-1.02cm}
    \includegraphics[width=0.36\linewidth]{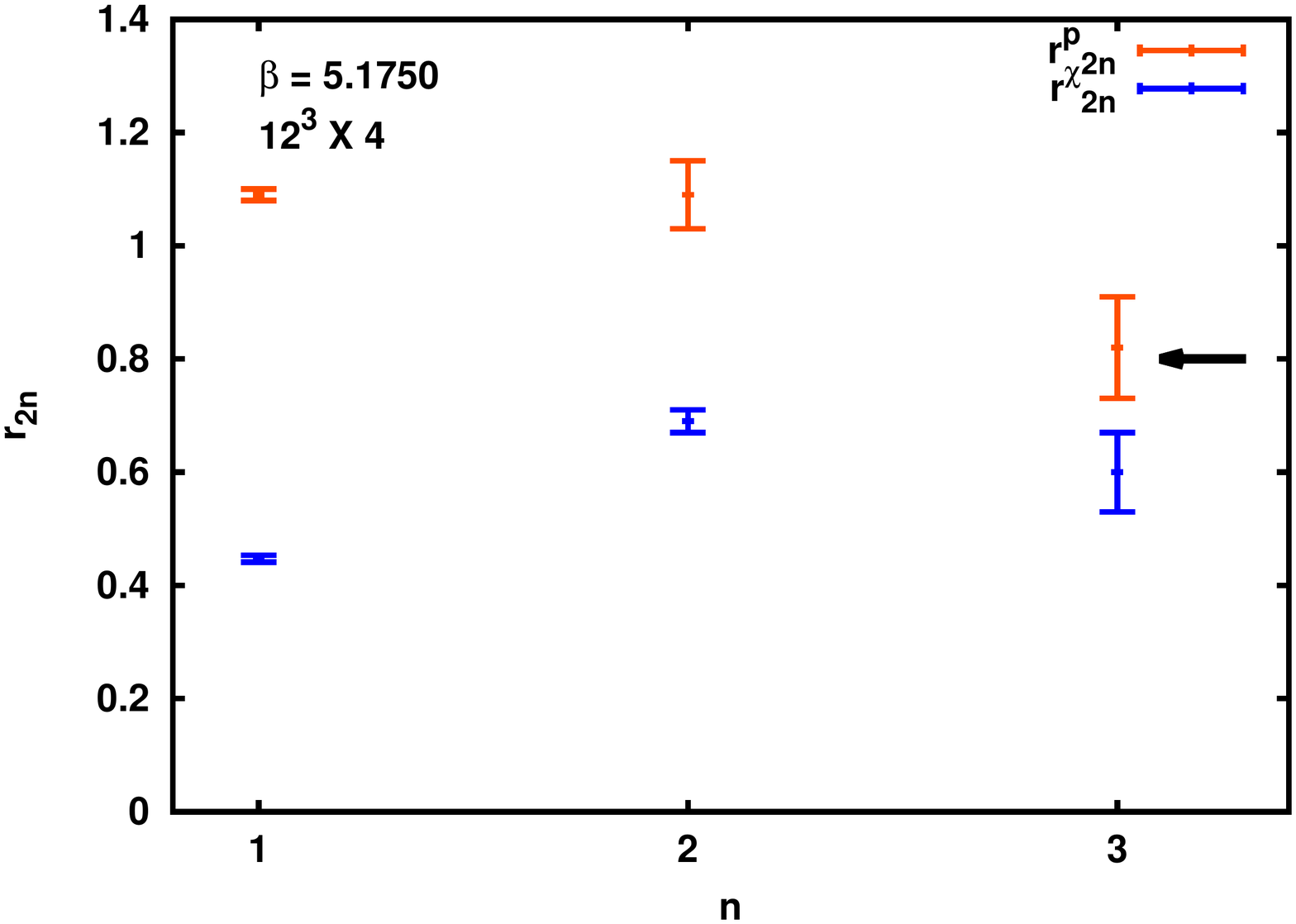}
\end{center}
    \vspace{-1.12cm}
    \caption{
        \label{fig:LY}
        Analysis of the Fisher zeros for the $N_t=4$ unimproved 
        staggered action. On the left is the imaginary part of the 
        closest Fisher zero as a function of the quark chemical 
        potential for the $12^3 \times 4$ lattice. In the center is 
        the infinite volume extrapolation for $\mu_q/T=0.8$. On the
        right are the ratio estimators for 
        the pressure $r^p_{2n}=\left| \frac{ (2n+2) (2n+1) \chi^B_{2n}}{\chi^B_{2n+2}} \right|^{1/2}$ and the 
        baryon number 
        susceptibility $r^\chi_{2n}=\left|\frac{2n(2n-1)\chi^B_{2n}}{\chi^B_{2n+2}} \right|^{1/2}$
        on a $12^3 \times 4$ lattice.
        One can see that the ratio estimators are in the same 
        ballpark as the critical endpoint estimate from the analysis
        of the Fisher zeros.
    }
\end{figure}

For coarse lattices, there exists practical algorithms to search for the Lee-Yang zeros, or the 
analogous zeros in gauge coupling parameter $\beta$, called the Fisher zeros $\beta_{LY}$. In 2004, such a 
study led to an estimate of the QCD critical end point on a course $N_t=4$ unimproved staggered lattice~\cite{Fodor:2001pe, Fodor:2004nz}.
We have repeated the old study, but with 50 times higher statistics and an exact algorithm and found a 
result very much consistent with the old one. The results for the Fisher zeros on a $12^3 \times 4$ lattice, the
infinite volume extrapolation at $\mu_q/T=0.8$  and the comparison
with the ratio estimators for the convergence radius can be seen in Fig.~\ref{fig:LY}. Here the convergence radius 
estimates from the ratio estimators are in the same ballpark as the critical point estimate from the analysis
of $\beta_{LY}$. 

The algorithms calculating the Fisher or Lee-Yang zeros are prohibitively expensive for more realistic lattices, 
there the only information available is the first few coefficients of the Taylor series.

\section{Ratio estimators for fine staggered lattices}
We know present our results for the ratio estimators for a fine $N_t=12$ 4stout-improved staggered lattice. 
The 4stout action was introduced in \cite{Bellwied:2015lba}. The details of the lattice calculation of the 
baryon fluctuations are explained in \cite{Borsanyi:2018grb}.
Here, we only give a brief description of some features relevant for our discussion. The calculation uses
simulations at imaginary chemical potentials ~\cite{deForcrand:2002hgr, DElia:2002tig}, where 
the value of the fluctuations $\chi^B_1, \chi^B_2, \chi^B_3$
and $\chi^B_4$ is calculated. 
We then fit to these values with a polynomial ansatz:
\begin{equation}
    p(T,\mu_B) = p_0(T) + \frac{1}{2!}  \chi^B_{2}(T)  \cdot \mu_B^{2} 
                        + \frac{1}{4!}  \chi^B_{4}(T)  \cdot \mu_B^{4}  + \dots 
                        + \frac{1}{10!} \chi^B_{10}(T) \cdot \mu_B^{10} 
\end{equation}
The fit is performed with a Bayesian procedure, where the higher order fluctuations $\chi^B_8$ and 
$\chi^B_{10}$ have a prior. This prior allows for the hadron resonance gas (HRG) prediction but does 
not prefer it, it also allows for the coefficients to grow faster than in the HRG, which could be
interpreted as a signal for critical behavior. More precisely, in the HRG we have $\chi^B_{2n+2}/\chi^B_{2n}=1$ and 
therefore the ratio estimator for the susceptibility $\chi^B_2$ grow 
as $r^\chi_{2n} \equiv \left|\frac{2n(2n-1)\chi^B_{2n}}{\chi^B_{2n+2}} \right|^{1/2} = \sqrt{2n(2n-1)}$. 
To have a finite radius of convergence on therefore needs the coefficients to grow like 
$\frac{\chi^B_{2n+2}}{\chi^B_{2n}} \sim n^2$. We instead see for the first few coefficients 
a result consistent with HRG.

\begin{figure}[h]
    \vspace{-1.1cm}
\begin{center}
    \includegraphics[width=0.69\linewidth]{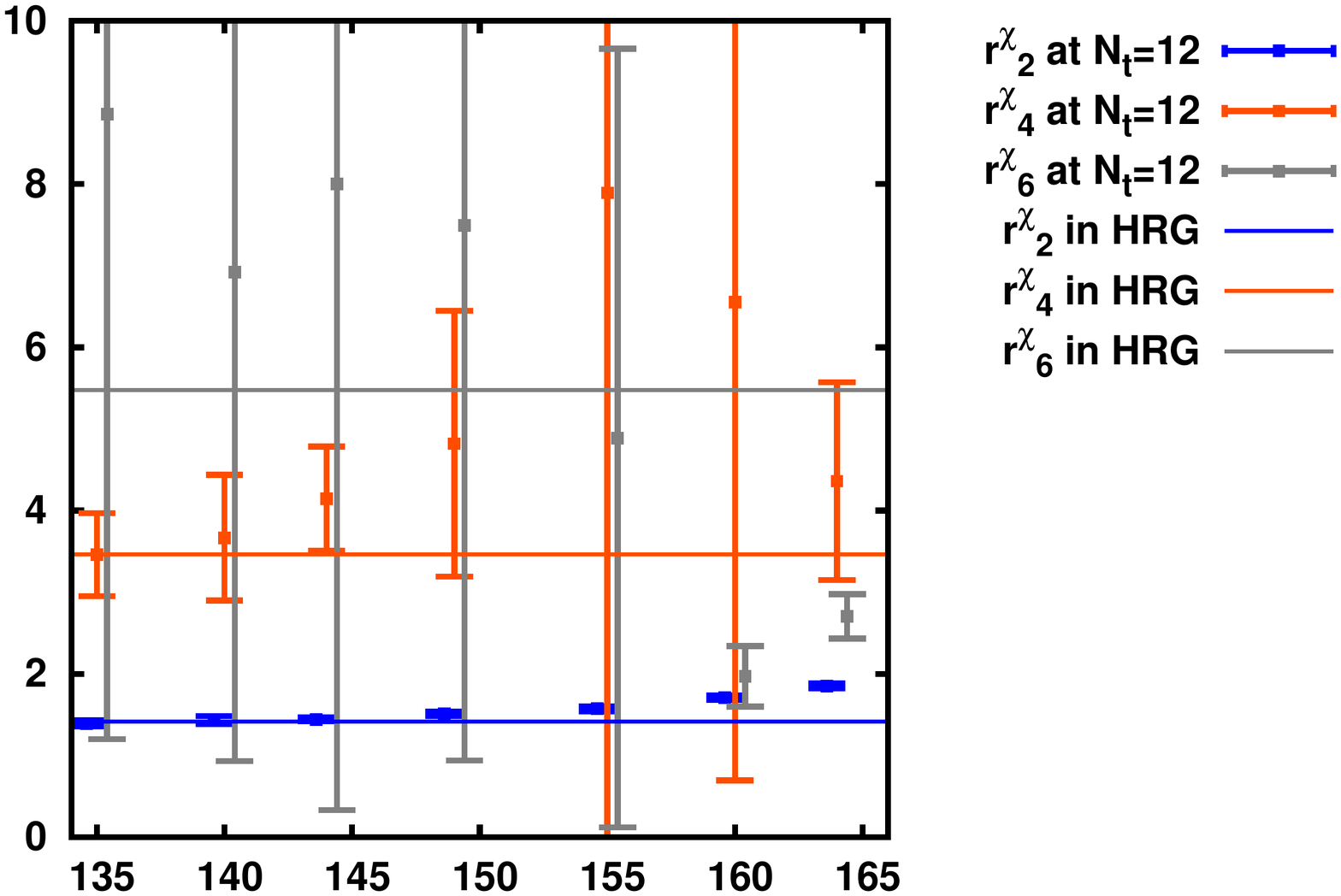}
\end{center}
    \vspace{-1.4cm}
    \caption{Ratio estimators of the susceptibility $r^\chi_{2n} \equiv \left|\frac{2n(2n-1)\chi^B_{2n}}{\chi^B_{2n+2}}\right|$ 
             on an $N_t=12$ 4stout-improved staggered lattice.}
    \vspace{-0.4cm}
\end{figure}

\section{A closer look at the lattice results for $N_t=12$}

To shed some light on the previous result consider the following toy model.
Start with some parametrization of the curve $\chi^{B}_1/\mu_B$ as a function of $T$ at $\mu=0$.
Assume that the only difference in the physics at finite $\mu$ is a shift in this curve in the 
$T$ direction. The inflection point of this curve is one possible definition of $T_c$, so
we shift the curve by using the curvature of the cross-over line found in the literature.
You now have a model prediction of $\chi^{B}_1$ for any finite $\mu$, and can differentiate it
a few times at $\mu=0$ to get estimates of $\chi^B_4$, $\chi^B_6$ and $\chi^B_8$. Comparison of this toy model with the actual lattice data is shown in Fig. 3.

\begin{figure}[h]
\begin{center}
    \includegraphics[width=0.58\linewidth]{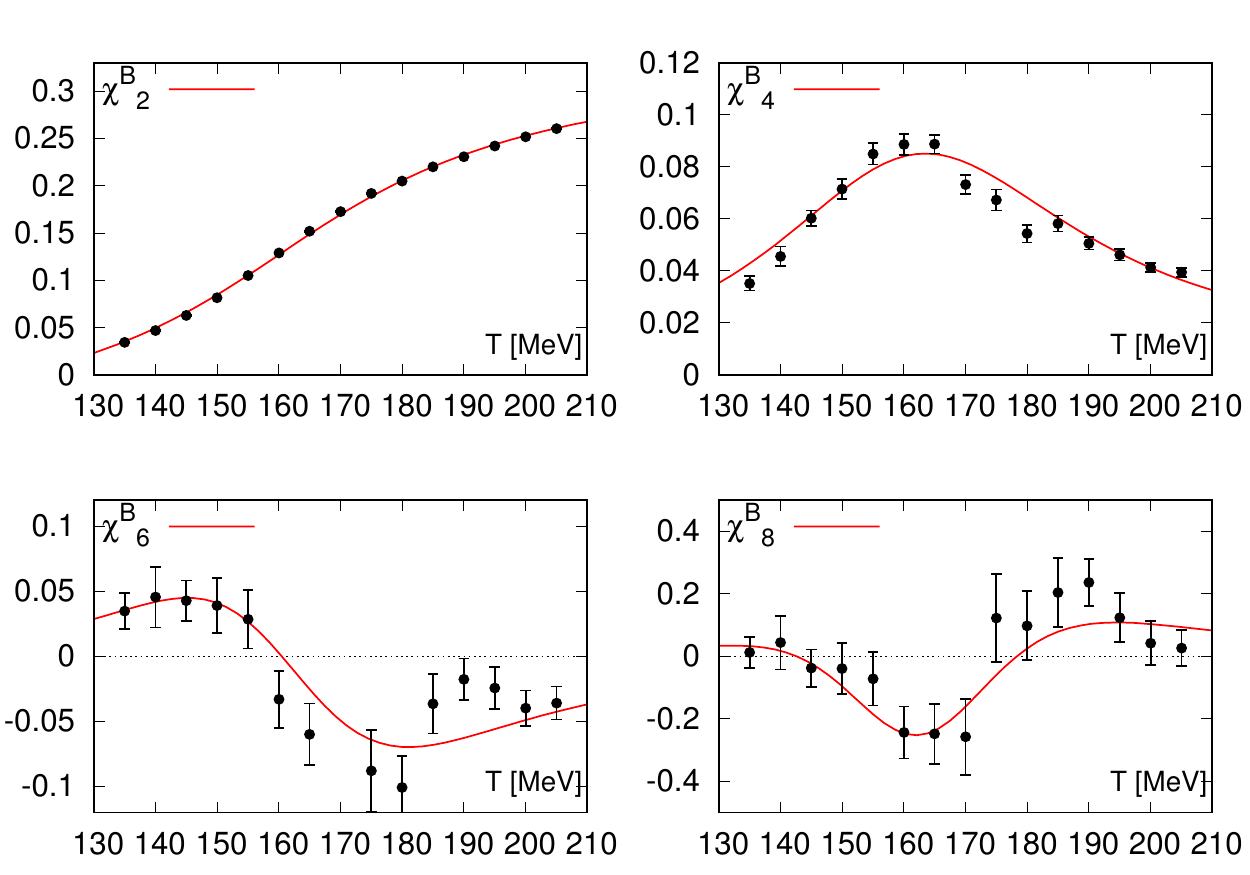}
\end{center}
    \vspace{-0.6cm}
    \caption{Comparison of the simple toy model described in the text with the actual lattices simulations.}
\end{figure}

The discussion of this toy model suggests the following non-trivial consequence for modeling:
In order to reproduce the lattice data on $\chi^B_2$, $\chi^B_4$, $\chi^B_6$and $\chi^B_8$ at the present accuracy all a model has to do is to
reproduce $\chi^B_2$ and the curvature of the cross-over line $\kappa$, without the crossover getting stronger
already at low values of the chemical potential. As long as these conditions are met, the features of the higher order baryon number susceptibilities will be automatically
reproduced. 

Among other things, this discussion suggests that near and above $T_c$, the curvature of the crossover 
implies $r^\chi_{2} < r^\chi_4$ and $r^\chi_4 > r^\chi_6$, a feature clearly visibly on the lattice data at $T=165\rm{MeV}$. Here 
the closest Lee-Yang zero most likely 
has a large imaginary part, and the ratio estimator is not expected to give a good estimate of the radius of
convergence.

As the existence of $\kappa$ implies $r^\chi_4 > r^\chi_6$ near or slightly
above $T_c$, while int the HRG $r^\chi_4 < r^\chi_6$, by continuity, there must
be a temperature $T_*<T_c$ where $r^\chi_4 = r^\chi_6$.
This is an apparent convergence in the first few ratio estimators, that 
however does not imply anything about criticality,
and when the statistical errors get small enough to see $T_*$ one should be careful not to misinterpret this apparent
convergence.
 
An other explicit example of a model reproducing the lattice data on baryon fluctuations, but having no 
critical point is found in \cite{Vovchenko:2017xad, Vovchenko:2017gkg}. 
Obviously, this does not 
necessarily mean that there is no critical point, just that if it exists then at the current levels of 
statistical uncertainty, the lattice results are not sensitive to it.



In summary, if a critical point is close to $\mu=0$, one may see it in a fast convergence of ratio estimators and 
this might be
what is happening for $N_t=4$. On the other hand, apparent convergence does not imply a critical point. In fact,
we argued that even in the case with no CEP, the ratios $r^\chi_4$ and $r^\chi_6$ will show 
apparent convergence somewhere below $T_c$. For our fine lattice ($N_t=12$ 4stout) the 
sign structure of $\chi^B_6$ and $\chi^B_8$ near $T_c$ is consistent with only a $\kappa$ and 
no criticality. At lower temperatures the data quickly become compatible with HRG, showing no traces
of criticality. Finally we note that none of 
these observations can be converted into a rigorous bound for 
the convergence radius of the Taylor series.
 
\appendix

\section*{Acknowledgements}
This project was funded by the DFG grant SFB/TR55. This work was supported by
the Hungarian National Research, Development and Innovation Office, NKFIH grants KKP126769 and
K113034. D.S. is supported by the DFG grant Heisenberg Programme (SE2466/1-2).
An award of computer time was provided by the INCITE program. This research used resources
of the Argonne Leadership Computing Facility, which is a DOE Office of Science User Facility supported
under Contract DE-AC02-06CH11357. The authors gratefully acknowledge the Gauss Centre for Supercomputing 
e.V. (www.gauss-centre.eu) for funding this project by providing computing time on the GCS
Supercomputer JUQUEEN at Julich Supercomputing Centre (JSC) as well as on HAZELHEN at HLRS 
Stuttgart, Germany. This material is based upon work supported by the National Science Foundation under 
grants no. PHY-1654219 and OAC-1531814 and by the U.S. Department of Energy, Office of Science,
Office of Nuclear Physics, within the framework of the Beam Energy Scan Theory (BEST) Topical Collaboration. 
C.R. also acknowledges the support from the Center of Advanced Computing and Data Systems at
the University of Houston.

\bibliographystyle{elsarticle-num}
\bibliography{<your-bib-database>}



\end{document}